\documentclass[12pt,dvips]{article}

\usepackage{amsmath,amssymb,amsbsy,amsfonts,hyperref,latexsym,graphicx,epsf}
\usepackage[round]{natbib} 
\usepackage{color}

\def\CC{{\rlap {\raise 0.4ex \hbox{$\scriptscriptstyle |$}} \hskip -0.2em C}}

\begin{document}

\bibliographystyle{apalike}



\vspace{0.5cm}

\noindent
{\bf \large Time-varying perturbations can distinguish among integrate-to-threshold models for perceptual decision-making in reaction time tasks.}

\bigskip

\noindent
{\bf Abbreviated title: Time-varying perturbations of decision-making models}

\vspace{0.5cm}

\noindent
{\bf Xiang Zhou$^{1*}$, KongFatt Wong-Lin$^{1* \dagger}$ and Philip Holmes$^{1, 2}$}

\vspace{0.25cm}

\noindent
$^{1}$Program in Applied and Computational Mathematics, \\
$^{2}$Department of Mechanical and Aerospace Engineering, \\
Princeton University,  Princeton, NJ 08544, USA \\

\vspace{0.15cm}

\noindent
$^{*}$ These authors contributed equally to this work.

\noindent
$^{\dagger}$ Corresponding author contact information: \\
KongFatt Wong-Lin. Phone: (609)-258-3685, fax: (609)-258-1735, \\
email: kfwong@math.princeton.edu

\vspace{0.15cm}
\begin{center}
{\Large Journal: }{\Large\it Neural Computation} 
\end{center}

\vspace{0.25cm}


\vspace{0.25cm}


\pagebreak


\begin{abstract}

Several integrate-to-threshold models with differing temporal
integration mechanisms have been proposed to describe the accumulation
of sensory evidence to a prescribed level prior to motor response in
perceptual decision-making tasks. An experiment and simulation studies
have shown that the introduction of time-varying perturbations during
integration may distinguish among some of these models. Here, we
present computer simulations and mathematical proofs that provide more
rigorous comparisons among one-dimensional stochastic differential
equation models. Using two perturbation protocols and focusing on the
resulting changes in the means and standard deviations of decision
times, we show that, for high signal-to-noise ratios, drift-diffusion
models with constant and time-varying drift rates can be
distinguished from Ornstein-Uhlenbeck processes, but not
necessarily from each other. The protocols can also distinguish
stable from unstable Ornstein-Uhlenbeck processes, and we show that a 
nonlinear integrator can be distinguished from these linear models by 
changes in standard deviations. The protocols can be implemented in 
behavioral experiments. 

\end{abstract}

\section{Introduction}
\label{s.intro}

Reaction time tasks have long been used to study human decision making
\citep{luce1986}. One paradigm requires subjects to detect or
discriminate sensory signals by making a  voluntary motor
response. Response time distributions obtained in such
perceptual-motor tasks allow inference of cognitive information
processing \citep{posner1978}. With the advent of methods for
recording neural activity in awake behaving animals, such tasks have
been adopted by neurophysiologists, especially in studies on monkeys
\citep{schall2003,GoldShadlen2007}, and neuronal firing rates in
several brain areas have been shown to correlate with motor
responses. The lateral intraparietal area, frontal eye fields, and
superior colliculus \citep{GoldShadlen2007} all exhibit activities
that ramp up over time toward a fixed level before a decision is
signaled (e.g. by a saccadic eye movement in the direction of the
recorded response field)
\citep{HanesSchall1996,roitman2002,churchland2008}. The slopes of the
ramps not only correlate with task difficulty (the harder the
task, the lower the slope), but also with response time (higher
slopes precede faster responses). These areas may therefore provide
neural substrates for integrating sensory information toward a
decision criterion before a perceptual decision is made.

Various integrate-to-threshold models have been proposed to describe
both response times and neurobiological mechanisms
\citep{RatcliffSmith2004}, including drift-diffusion models
\citep{Ratcliff1978,Mazurek2003,RatcliffSmith2004,Ditterich2006c,Simen2006,Ratcliff2008} 
and attractor neural networks with mutual inhibition 
\citep{BrownHolmes2001,UsherMcClelland2001,Wang2002,WongWang2006,LoWang2006}.
All share a common mechanism: accumulation or integration of
sensory inputs toward a prescribed threshold, the first crossing of
which determines the decision and response time. These models may
nonetheless be distinguished by the details of the integration
process: drift-diffusion models typically accumulate evidence at a
constant rate like a biased random walk, while attractor networks have
unstable or stable steady-states  that can, respectively, accelerate
or decelerate the integration process
\citep{UsherMcClelland2001,WongWang2006}.

Often only behavioral data are collected for human subjects, and
fits based on response times and choice accuracies are sometimes
unable to distinguish among competing models
\citep{RatcliffPR1999,Ratcliff2006}. Moreover, while cellular
recordings in awake behaving animals offer direct insights into the
integration process, fitting of both behavioral and neural data may
still not suffice, especially when models incorporate multiple
features and depend on multiple parameters. For example, a
drift-diffusion model with a time-dependent `urgency' signal
\citep{Ditterich2006c,churchland2008} may be
difficult to separate from a recurrent network model with strong
self-excitation \citep{Wang2002,WongWang2006}. Thus far, few 
principled attempts have been made to tease apart different integration
mechanisms. 

Subthreshold electrical microstimulation of neural activities in 
behaving animals may provide more conclusive tests 
\citep{CohenNewsome2004}. 
In \citet{Ditterich2003}, microstimulation of sensory neurons was shown
to affect the speed of the decision. More interestingly, in
\citet{Hanks2006}, stimulation of cells with choice targets in the
recorded neuronal response fields speeded decisions, but also
\emph{reduced} decision speeds when choice target directions were
opposite to the recorded response fields, thus providing evidence of
mutual inhibition. Perturbations need not be invasive, and so can
be used in human studies: e.g., \citet{HukShadlen2005} employed a
brief motion pulse in the background of the primary visual stimulus; 
although it neither determined the choice nor influenced 
rewards, the pulse had a significant effect on response times. 

Earlier modeling efforts \citep{HukShadlen2005,Wong2007,WongHuk08} have
addressed the data of \citet{HukShadlen2005}, but they employed  many
parameters and direct comparisons were not made between models.  Here,
we conduct a more rigorous study using simpler  models, and seek more
objective comparisons among them.  We approximate four neural firing
rate models as linear, scalar, stochastic differential equations
(SDE), and we ask how their predicted response times are affected by
short piecewise-constant perturbations with varying onset times and
amplitudes. By  comparing changes in means and standard deviations of
response times, we demonstrate that the perturbations suffice to
distinguish among the models. Finally, we show that a nonlinear
integrator model, qualitatively similar to that of
\citet{WongWang2006}, behaves much like one of the linear models.

\section{Methods}
\label{s.methods}

\subsection{Reduction to one-dimensional linear integrate-to-threshold models}
\label{ss.redn}

Two-alternative forced-choice decision processes can, in essence, be
modeled by two populations of excitatory neurons each endowed with
self-excitatory  connections and mutual inhibition via a shared
inhibitory population. Each of the three populations can then be
represented in a coarse-grained firing rate model by a single unit
whose state describes the population-averaged activity $r_j(t)$  of
the corresponding neuronal pool
\citep{Wilson-Cowan72,Wilson-Cowan73}. Here we review the further
reduction of a firing rate model, under suitable hypotheses, to a
one-dimensional dynamical  system. For additional details, see
\citet{brown2005,Bogacz2006}. 

Consider first the deterministic equations describing the
firing rates of two excitatory ($E$) populations and a common
inhibitory ($I$) population:
\begin{eqnarray}
\tau_{E} \frac{dr_{1}}{dt} &=& -r_{1} + F_{E}(r_{1},r_{2},r_{I},I_{1}) ,
 \nonumber \\
\tau_{E} \frac{dr_{2}}{dt} &=& -r_{2} + F_{E}(r_{2},r_{1},r_{I},I_{2}) ,
 \label{eq:firingrate1} \\
\tau_{I} \frac{dr_{I}}{dt} &=& -r_{I} + F_{I}(r_{1},r_{2}) .
\nonumber 
\end{eqnarray}

Here $\tau_{E}$ and $\tau_{I}$ are the synaptic time constants for the
$E$ and  $I$ units, and $r_{1}$, $r_{2}$ and $r_{I}$ are respectively
the activity of neural  units $1$, $2$ and $I$. The excitatory
populations $1$ and $2$ are selective to stimuli $1$ and $2$,
respectively. $F_{E}$ and $F_{I}$ are their input-output functions, and
the overall input to each unit, with $i=E$ or $I$, is $I_{i}
=I_{recurrent,i} + I_{stimulus,i}$. The decision time is the first passage
time from stimulus onset to the first of $r_1$ or $r_2$ reaching a
prescribed decision threshold, which thereby signals choice $1$ or
$2$. Since non-decision latencies  (e.g. signal transduction and motor
preparation) are usually assumed to be independent of stimulus
strength $I_{stimulus}$, we shall model the response time as the
decision time plus a constant latency. Henceforth we use the phrases
`decision time', `reaction time' and `response time' interchangeably.

If the decision dynamics passes near a saddle point and moves
along its unstable manifold \citep{BrownHolmes2001,Bogacz2006},
the functions $F_{E}, \, F_{I}$ may be linearized such that
Eq.~\ref{eq:firingrate1} simplifies to:
\begin{eqnarray}
\tau_{E} \frac{dr_{1}}{dt} &=& -r_{1} + (\alpha r_{1} - \beta r_{I}
 + I_{1} + I_{0, E}) , \nonumber\\
\tau_{E} \frac{dr_{2}}{dt} &=& -r_{2} + (\alpha r_{2} - \beta r_{I}
 + I_{2} + I_{0, E}) , \label{eq:firingrate2} \\
\tau_{I} \frac{dr_{I}}{dt} &=& -r_{I} + (\gamma r_{1} + \gamma r_{2}
 + I_{0, I}) . \nonumber
\end{eqnarray}
Here $\alpha, \beta, \gamma$ are the recurrent synaptic coupling
strengths for  self-excitation, inhibitory-to-excitatory, and
excitatory-to-inhibitory connections, and $I_{0, E}$ ($I_{0, I}$) is
the constant  background input to all the excitatory (inhibitory)
cells from outside  the local circuit. Unlike \citet{Wang2002}, we
exclude excitatory connections between $r_{1}$ and $r_{2}$ and
self-inhibitory connections, retaining only essential features.  If
$\tau_{I} \ll \tau_{E}$, we can further assume that the relatively
fast dynamics of $r_{I}$ equilibrates rapidly, such that  $r_{I}
\approx \gamma (r_{1}+r_{2}) + I_{0, I}$, and Eq.~\ref{eq:firingrate2}
becomes
\begin{eqnarray}
\tau_{E} \frac{dr_{1}}{dt} &=& -r_{1} + (\alpha-\beta \gamma) r_{1}
 - \beta \gamma r_{2} + I_{1}  + (I_{0, E} - \beta I_{0, I}) , \label{eq:fr1} \\
\tau_{E} \frac{dr_{2}}{dt} &=& -r_{2} + (\alpha-\beta \gamma) r_{2}
 - \beta \gamma r_{1} + I_{2} + (I_{0, E} - \beta I_{0, I}) . \label{eq:fr2}
\end{eqnarray}
Defining a new variable $X \equiv r_{1} - r_{2}$ and subtracting 
Eq.~\ref{eq:fr2} from \ref{eq:fr1}, we obtain
\begin{equation}
\frac{dX}{dt} = k X +  (I_{1}(t)-I_{2}(t))/\tau_{E},  \nonumber
\label{eq:subtracteqn}
\end{equation}
where $k \equiv (\alpha - 1)/\tau_{E}$ contains the excitatory
coupling strength and leak, and  $(I_{1}-I_{2})/\tau_{E}$ is
proportional to the difference in inputs. The background inputs
$I_{0, E}$  and $I_{0, I}$ and coupling strengths $\beta$ and
$\gamma$ cancel out.

\begin{figure}[htb!]
 \centering
 \includegraphics[scale=0.44]{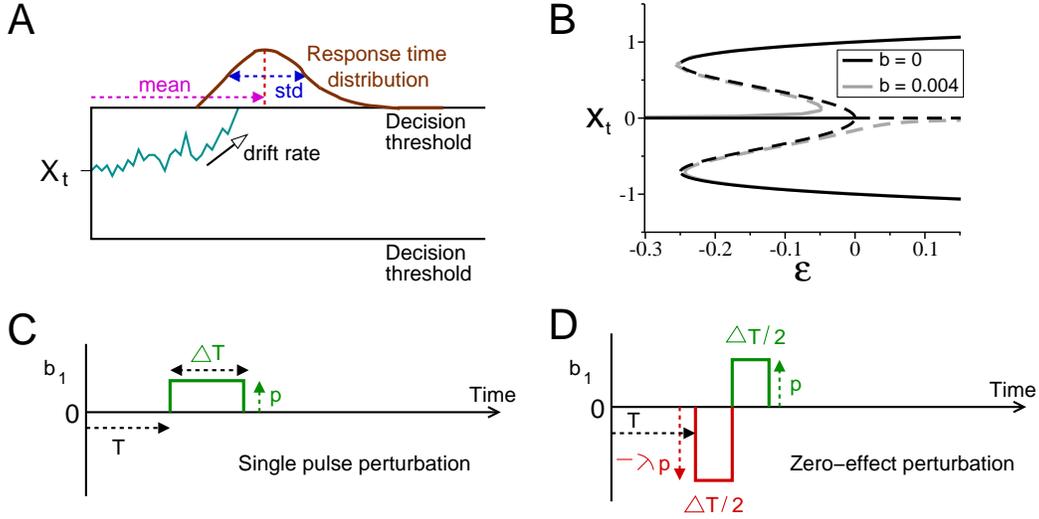}
\caption{One-dimensional integrate-to-threshold models and
perturbation protocols.  A: a general model for a two-alternative
forced-choice reaction time task. Here the noisy state $X_t$, which
quantifies the difference in firing rates between two competing
excitatory cell populations, reaches the upper decision threshold in a
sample trial, and choice is correct since threshold is in same
direction  as drift rate. Multiple sampling provides the response time
distribution.  B: the SPB model with unbiased input exhibits a
subcritical pitchfork bifurcation at $\varepsilon = 0$.
Black curves show branches of stable (solid) and unstable
(dashed) fixed points in the noiseless limit; branches for biased 
input shown in gray. C: a short pulse can be repeated over a
range of onset times $T$ after primary stimulus onset at $t=0$ to test
effects on decision times. D: a pulse-antipulse perturbation whose
leading pulse amplitude (red) can be adjusted to determine when
decision times are unaffected by the combined pulse,  yielding a
zero-effect perturbation (ZEP). See text for further details. }
\label{fig:scheme}
\end{figure}

Generalizing the inputs $I_{1}$, $I_{2}$ to be time-varying and
including additive noise, the reduced dynamics  is described by a
one-dimensional SDE of the form
\begin{equation}
d X_{t} = b(X_{t},t) dt + \sigma dW_{t} ,
\label{eq:general}
\end{equation}
where $b(X_{t},t) \equiv k X_{t} + (I_{1}(t)-I_{2}(t))/\tau_{E}$ and $\sigma$ 
is the standard deviation of the noise, which is assumed to be a Wiener
process with increments $dW_{t}$ drawn from a normal distribution with 
zero mean and unit variance. Eq.~\ref{eq:general} provides a general
description for noisy accumulator models, as illustrated schematically
in Fig.~\ref{fig:scheme}A. 

\subsection{Linear integrate-to-threshold models}
\label{ss.linmods}

We are now in a position to introduce four linear instances of
Eq.~\ref{eq:general}. When $k < 0$ (e.g. self-excitation $\alpha$ is
relatively weak), we have a classical Ornstein-Uhlenbeck (OU) process
\citep{Uhlenbeck-Ornstein,Wang-Uhlenbeck}. If the decision threshold
does not intervene, solutions eventually decelerate and approach
a stable steady-state $X_{steady-state} =
(I_{1}(t)-I_{2}(t))/(|k| \tau_{E})$. We denote this stable OU process
SOU; in contrast, if  $k >0$  (stronger self-excitation),
sample paths accelerate away from a fixed point in an unstable OU
process, denoted UOU. SOU and UOU provide reduced descriptions of the
leaky competing accumulator of \citet{UsherMcClelland2001} that
respectively produce recency and primacy effects. 

In the special case of $\alpha = 1$ (i.e. $k = 0$) and constant
stimuli $I_{1}$, $I_{2}$,  Eq.~\ref{eq:general} becomes a pure
drift-diffusion (CD) equation \citep{Ratcliff1978}
\begin{equation}
d X_{t} = b_{0} dt + \sigma dW_{t} ,
\label{eq:CD}
\end{equation}
in which  the constant drift rate $b = b_{0} \equiv
(I_{1}-I_{2})/\tau_{E}$ is proportional to the difference in stimuli. This
CD model is similar to, but differs in detail from, the balance 
between leakage and inhibition in \citet{Bogacz2006}.)

In some variants of the CD model, drift rates can vary with time. 
This could be due to dynamic stimuli, to some form 
of `urgency' induced by the task design \citep{Ditterich2006c,Ditterich06b,churchland2008}, 
to fluctuating attention \citep{Smith2004} or to other `top-down' effects 
\citep{LiuHCoh08}. This time-dependent (TD) diffusion model is described by:
\begin{equation}
d X_{t} = b(t) dt + \sigma dW_{t} .
\label{eq:TD}
\end{equation}
We focus on the simple case $b(t) = b_{0} t$, since similar results
obtain for other functions of time, but in Appendix B, we treat a
modified TD model with time-varying perturbation amplitude.

\subsection{A nonlinear integrate-to-threshold model}
\label{ss.SPB}

The final model to be considered is qualitatively
similar to that of \citet{Wang2002,WongWang2006}.  
It captures stochastic dynamics in the neighborhood of a subcritical
pitchfork bifurcation \citep{GH83,BrownHolmes2001,Strogatz}
that occurs as the stimulus input level is varied 
\citep{RoxinLedberg08,WongWang2006}:
\begin{equation}
\tau_X dX_{t} = [\varepsilon X_{t} + X_{t}^{3} -X_{t}^{5} + b(t)] dt + \sigma dW_{t} .
\label{eq:normalform}
\end{equation}
Here $b(t) \equiv b_{0}$ (constant) and $\varepsilon$ represent the
biased and non-biased stimulus inputs respectively
(Fig.~\ref{fig:scheme}B). $\tau_X$ controls the overall temporal dynamics. 
Eq.~\ref{eq:normalform} may be derived by normal form theory 
\citep{GH83,RoxinLedberg08}: see 
Fig.~\ref{fig:scheme}B for an illustration of the branches of stable
and unstable fixed points in the noise-free limit. If $|\varepsilon|$
and $|b|$ are  sufficiently small, this model without the $X_{t}^5$
term approaches that of \citet{Wang2002,WongWang2006,RoxinLedberg08}. 
We include $-X_{t}^5$ to prevent unrealistic runaway activity, and, for
simplicity, first assume a symmetrical system with $b_0=0$, and then
consider biased stimuli $ \vert b_0 \vert > 0$. This model is denoted
SPB.

\subsection{Perturbation protocols}
\label{ss.perts}

\subsubsection{Single pulse perturbation with varying onset time}

We consider additive perturbations $b_1(t)$, under which 
Eq.~\ref{eq:general} becomes 
$dX_{t} = [b(X_{t},t) + b_1(t)] dt + \sigma dW_{t}$, 
where $b_1(t)$ is applied from time $T$ to $T+\Delta T$ and
$b_1(t) \equiv 0$ otherwise. We employ two piecewise-constant
forms, the first being the step function used by \citet{HukShadlen2005}:
\begin{equation}
b_{1}(t) = \left\{
\begin{array}{ll }
0   , & t \leq T , \\
p   ,  & T< t \leq  T+\Delta T , \\
0   , & t > T+\Delta T ,
\end{array}\right.
\label{eq:singlepulse}
\end{equation}
in which the amplitude $p$ can be positive or negative, 
assisting or opposing the unperturbed drift rate $b_0$ due to 
the `primary' stimulus. We fix the duration $\Delta T$ at $10\%$ of
the unperturbed  mean first passage time $\tau_{0} / 10$, apply
$b_{1}$ at different onset times $T$ and ask how the mean and
standard deviation of the decision time change.

\subsubsection{Zero-effect pulse-antipulse perturbation}

The second protocol uses a double pulse of the form 
\begin{equation}
b_{1}(t) = \left\{
\begin{array}{ll }
0   , & t \leq T , \\
- \lambda p   ,  & T< t \leq  T+\Delta T /2 , \\
p  , & T + \Delta T/2 \leq t < T + \Delta T, \\
0   , & t > T+\Delta T ,
\end{array}\right.
\label{eq:ZEP-lambda}
\end{equation}
where $\lambda$ is the relative height of the first pulse to the second. 
The second opposite-signed pulse 
attempts to reduce or cancel the expected change due to the first, and
if the overall perturbation $b_1$ leaves the mean first passage time
unchanged, we call it a \emph{zero-effect perturbation} or ZEP. 
This protocol, a variant of the paired pulse suggested in
\citet{Wong2007}, is only applied early in the integration process to
avoid interference from the decision threshold.  We shall seek
critical values of $\lambda$ for which ZEPs occur for each
model. Figs.~\ref{fig:scheme}C and D illustrate both protocols.

\subsection{Simulations and parameter values}
\label{ss.sim}

Except for the SPB model, we first consider integration to a single
decision threshold in the direction of the drift. This reduces the 
number of parameters, simplifies mathematical analyses, 
and helps isolate key effects. It applies to 
easy tasks in which drift rates are (relatively) high and errors
rare. We subsequently relax this condition in our simulations.
For the SPB model we first set $I_{1}=I_{2}$, corresponding to difficult 
tasks and high error rates, and employ two thresholds. We then consider 
$I_{1} \neq I_{2}$ and a single threshold, representing easy tasks.

Signal-to-noise ratios were chosen such that the variance in first
passage (decision) times is significant, but not so great that an
unreasonably large number of trials is needed to average out the
noise, and we selected perturbation amplitudes, durations
and onset times such that both protocols cause small but significant
effects (e.g. maximum changes of $\sim 10\%$ in mean decision
times). In particular, durations were an order of magnitude smaller
than mean decision times $\tau_0$, as in \citet{HukShadlen2005}. For the 
first protocol, onset times $T$ were varied from stimulus onset at $t = 0$ 
until there were no significant effects on mean decision time. In the
second protocol we require $T \ll \tau_0$. We set $X_0 = 0$ at
$t=0$ to represent unbiased initial conditions, and the remaining
parameters are chosen to ensure `realistic' behavior (e.g. the stable
fixed point is above threshold for SOU; the unstable fixed point
is below $X_0 = 0$ for UOU). Table~\ref{tab.1} lists parameter
values.

\begin{table}[htb]
\begin{center} 
\begin{tabular}{|c|c|c|c|c|c|}
  \hline
  Model & $b(X_{t}, t)$ & Parameters &  $\zeta$ & $\Delta T$ & $p$ \\
  \hline\hline
  CD  & $b_0 + b_1$ & $b_0 = 5, \sigma = 2.449$ & $20$ & $0.4$ & $5$   \\
    \hline
  TD  & $b_0 t + b_1$ & $b_0 = 4, \sigma = 2.828$ &$20$ & $0.1$ & $4$  \\
    \hline
  SOU & $k X_t + b_0$ & $k = -1, b_0 = 8$,  & $7$ & $0.4$ & $2$ \\
  & $+b_1$ & $\sigma = 1.414$ & & & \\
    \hline
  UOU & $k X_t + b_0$ & $k = 0.2, b_0 = 5$, & $20$ & $1$ & $2$  \\
  & $+b_1$ & $\sigma = 1.414$ & & & \\
    \hline  SPB &  $\varepsilon X_t + X_{t}^{3}$ & $\varepsilon = -0.3, 
    0.05, b_0 = 0$  & $\pm 0.75$ & $5$ & $0.005$ \\
 & $- X_{t}^{5}+b_0$ & or $0.004$, $\sigma = 0.01$, & & & or $0.0008$\\
 & $+ b_1$& $\tau_X = 20$ & & & ($b_0=0.004$) \\
  \hline
\end{tabular}
\end{center}
\caption{Parameters used in simulations. Drift-diffusion model
with constant drift rate (CD), and with time-varying drift rate (TD); 
SOU (UOU) stable (unstable) Ornstein-Uhlenbeck processes; 
SPB: nonlinear model. See Eqs.~\ref{eq:general}-\ref{eq:normalform} and 
protocols~\ref{eq:singlepulse} and \ref{eq:ZEP-lambda} for 
details. Here $\zeta$ is the distance from starting point
$X=0$ to decision threshold, and for SPB $\varepsilon = -0.3$ 
before stimulus appears at $t = 0$ and $\varepsilon = 0.05$
for $t \geq 0$. Perturbation durations $\Delta T$ and amplitudes $p$
are values used in the first perturbation protocol. Parameters for
the linear models with smaller signal-to-noise 
ratios are specified in Section \ref{ss.SNR}.}
\label{tab.1}
\end{table}

We use a  forward Euler-Maruyama scheme
\citep{Higham01}, integrating a  sample path until it hits the
prescribed decision threshold and recording the corresponding first
passage time, at which the trial ends and the next begins. After
collecting an appropriately large ensemble, we extract the first and
second moments of the first passage time from the
samples. The simulation is run once for each one of the different
perturbations. For the models CD,  TD, SOU and UOU, the step size
is $10^{-3}$ of a time unit and the sample size is $N = 10^{6}$, so that
errors in the moments are of order $1/\sqrt{N} \approx 10^{-3}$. Since
changes in moments due to the perturbations are of order $10^{-2}$,
these choices of time step and sample size reliably capture the
effects of perturbation. The mean exit time for the
unperturbed SPB model is $\sim 80$, so that a time step of $0.01$
and a sample size is $10^{6}$ suffices. We ran noisy simulations
of this model with $\varepsilon = -0.3$ for $t<0$ to represent
pre-stimulus activity,  and then switched to $\varepsilon = 0.05$ for
$t \geq 0$ so that $X = 0$ undergoes a pitchfork bifurcation and
becomes unstable, thus forcing a choice (see Fig.~\ref{fig:scheme}B. 
Magnitudes of $\varepsilon$ and $\sigma$ for SPB are much smaller 
than the corresponding $b_0$ and $\sigma$ for the linear models 
because the $+ X_{t}^{3}$ term accelerates solutions toward the thresholds). 

The parameters are not independent, and their number can be reduced by
one in all models by dividing Eq.~\ref{eq:general} by $b_0$, and rescaling
$\sigma \rightarrow \sigma / b_{0}$, $p \rightarrow p / b_{0}$,
$\zeta \rightarrow \zeta / b_{0}$,  $k \rightarrow k / b_{0}$ and the
dynamical variable $X_{t} \rightarrow X_{t} / b_{0}$.  The resulting
CD and TD models are described by $2$  parameters while the SOU and
UOU  models require 3 and the SPB model $4$. Two additional
parameters describe the amplitude $p$ and  duration $\Delta T$ of the
perturbation.

We focus on qualitative patterns of changes in response times
and, to make unbiased comparisons, we adopt a
dimensionless measure of the changes in means and 
standard deviations by normalizing the relative
changes with respect to their unperturbed values.

\section{Results}
\label{s.results}

\subsection{Single pulse perturbations of linear models}
\label{ss.spplin}

\begin{figure}[htb!]
 \centering
  \includegraphics[scale=0.6]{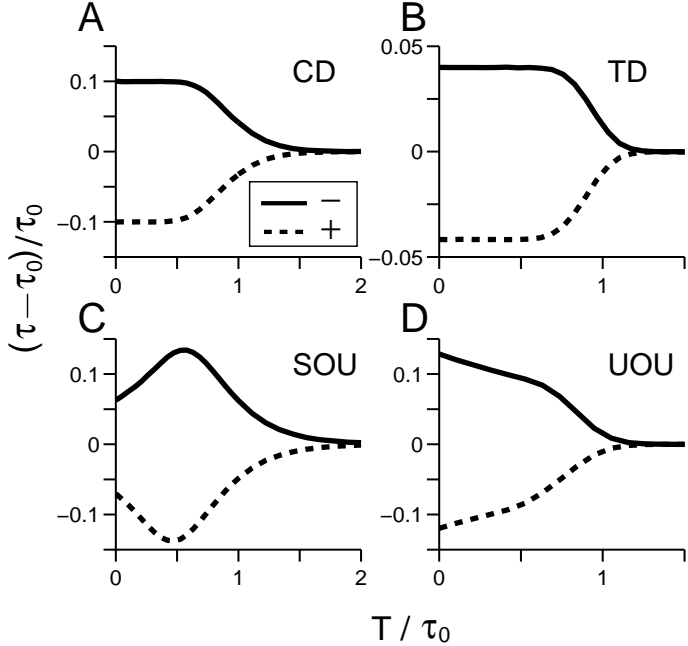}
\caption{Changes in mean decision times under single pulse
perturbations with varying onset times superimposed on positive
drift rates for linear integrators CD, TD, SOU, and UOU
(panels A-D).  Vertical axes: normalized changes in
mean first passage times $(\tau - \tau_{0}) / \tau_{0}$; horizontal axes:
normalized perturbation onset times $T / \tau_{0}$. Changes due to
negative pulses shown solid ($-$); changes due to positive pulses
dashed ($+$). CD and TD show near-constant changes for
$T / \tau_{0} \le 0.5$ that are substantially greater for CD (note vertical
scale on panel B),  SOU exhibits a maximum at $T / \tau_{0} \approx
0.5$ (``optimal'' perturbation), and UOU shows monotonic
decrease. Effects decrease in all cases as $T$ 
approaches and passes $\tau_{0}$ due to thresholding (see
text).}
\label{fig:T_mean}
\end{figure}

Using a single pulse perturbation with varying onset time $T$, we
estimated changes in mean decision time for the four linear
integrate-to-threshold models as described above, obtaining the
results shown in  Fig.~\ref{fig:T_mean}. Imposed early during the
integration process, a brief pulse in the same (opposite) direction
as the drift rate significantly decreases (increases) mean decision
times in all four models, as shown by the dashed (solid)
curves. These effects fade as $T$ increases due to a 
\emph{thresholding effect}: for later perturbations, 
more trials have already crossed threshold, and so 
there are relatively fewer trials that are perturbed than 
unperturbed. Thus, when averaging over all trials, 
the influence of perturbations is progressively reduced. 
Note that unlike the work of \citet{HukShadlen2005,Wong2007}, 
trials that have crossed threshold before and during perturbation 
are not excluded in the averaging process. This helps to 
reduce noise in the data, especially with late perturbation 
onset times. 

In Appendix A, we prove that this basic pattern -- positive 
pulses advance mean decision times and negative pulses retard
them -- must hold provided that the perturbation occurs sufficiently
early. The proof applies to a broad class of nonlinear systems
including the SPB model of Eq.~\ref{eq:normalform}, provided that
$\varepsilon > 0$ and thresholds lie inside the region in which
drift magnitude increases with $X$ (i.e., well below the activation 
levels of the stable fixed points).

Note that the drift-diffusion models (CD, TD) and the OU models (SOU, UOU) 
respond distinctly to early perturbations: the former exhibiting almost
constant changes in mean decision times (panels A and B), and
the latter respectively showing increasing and decreasing effects
(panels C and D).  Moreover, due to thresholding, SOU is alone in
having an onset time for which the perturbative effect is maximized.

\begin{figure}[htb]
 \centering
  \includegraphics[scale=0.6]{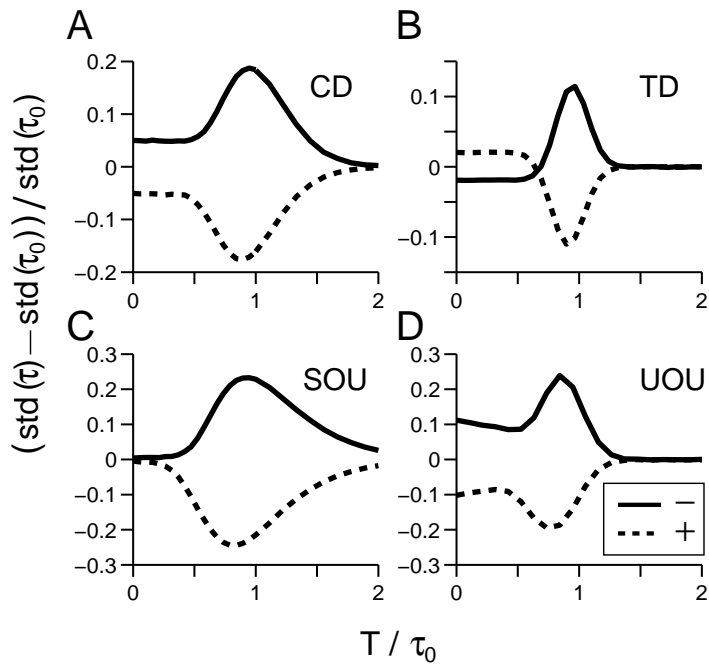}
\caption{Changes in standard deviations of decision times under single
pulse perturbations with varying onset times. All cases exhibit a maximum in
normalized standard deviation (std$(\tau)-$std$(\tau_{0})) / $std$(\tau_{0})$,
but only UOU has a (shallow) minimum. Effects of early pulses on
all four integrators are distinct, as described in text. Format and conventions
are as in Fig.~\ref{fig:T_mean}. }
\label{fig:T_std}
\end{figure}

We next investigate how standard deviations are affected  by
perturbation onset time: see Fig~\ref{fig:T_std}. Richer patterns
arise than those of Fig.~\ref{fig:T_mean}. Both CD and TD exhibit
similar near-constant changes for early onsets, but the
\emph{directions} of changes for TD are opposite to those for CD,
and the curves cross before reaching maxima.
Standard deviations for SOU have a similar pattern to its mean, 
but UOU exhibits initial decreases followed by increases to a peak, 
unlike its mean. Optimal onset times exist for which the
four standard deviations are maximally affected, 
and all four cases show distinct patterns. 

In the TD model considered above we assumed that the
perturbation is not affected by the time-dependent gain $b(t)$.
This may not hold if the perturbation enters via the same sensory
pathway as the stimuli. In Appendix B we consider a case 
in which perturbation and drift are affected in the same way. We
show that this modified TD model differs qualitatively only for
early perturbation onset, for which it yields reduced effects.

\subsection{Zero-effect perturbations of linear models}
\label{ss.zeplin}

We now consider the zero-effect protocol of Eq.~\ref{eq:ZEP-lambda},
appealing to the fact that, in the noise-free case, the
first passage time is a monotone function of the ratio $\lambda$ of
the amplitudes of the opposing pulses. The critical $\lambda =
\lambda^*$ that determines a ZEP is therefore given by the unique zero of
\begin{equation}
\tau(\lambda) = \tau_0 ,
\label{eq:ZEP-lambda*}
\end{equation}
where $\tau(\lambda)$ describes the functional dependence of
passage time on $\lambda$.
Mean first passage times appear to remain monotone with respect to
$\lambda$ when noise is included. Asymptotic 
methods can be used to approximate passage time
moments for small perturbations \citep{lindner2004}, but the lack of
compact formulae for passage time distributions (PDFs) in most cases,
including OU processes, and the need for double integrations
\citep{lindner2004} makes this approach generally intractable.

However, ZEPs may be analytically approximated for early
onset times $T$ with $T' := T+\Delta T \ll \tau_0$, such that the
probability of threshold crossing prior to $T'$ is negligible. In this
case a sufficient condition is that the PDFs of the perturbed and
unperturbed processes are identical at $T'$, since for $t > T'$, both
processes are governed by the same drift and noise. They are therefore
indistinguishable in the limit $T, \Delta T \rightarrow 0$. Before
deriving explicit expressions for $\lambda^*$ under this assumption
we present the results of numerical simulations with an intuitive
explanation.

\begin{figure}[htb!]
 \centering
 \includegraphics[scale=0.625]{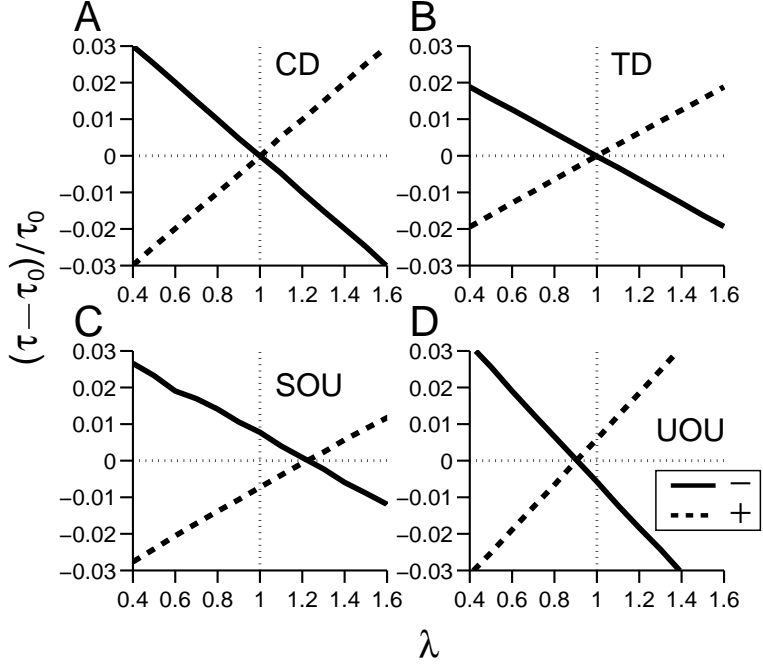}
\caption{Zero-effect perturbations of linear integrate-to-threshold
models CD, TD, SOU, UOU (panels A-D). Vertical axes: normalized change
in mean first passage times $(\tau - \tau_{0}) / \tau_{0}$; horizontal
axes: relative amplitude $\lambda$ of first pulse compared to
second. Zeroes of solid lines ($-$, positive pulse followed by
negative pulse, $p < 0$) and dashed lines ($+$, negative pulse
followed by positive pulse, $p > 0$) identify ZEPs at $\lambda^*
\approx 0.9987, 1.0039$ (CD); $0.9952, 1.0004$ (TD); $1.2271, 1.2230$
(SOU) and $0.9066, 0.9034$ (UOU). }
\label{fig:ZEP-lambda_mean}
\end{figure}

Since the unperturbed SDE \eqref{eq:general} is independent of the
current state $X_t$ in the CD and TD cases, we expect that
antisymmetric pulses with $\lambda^* = 1$, in which the pulses
precisely cancel, will produce ZEPs. In contrast, responses to inputs
decay with time for SOU, implying that the first pulse must be larger
than the second for their net effect to cancel at perturbation offset
$t = T'$ ($\lambda^* > 1$). For UOU the reverse should hold ($\lambda^*
< 1$). Fig.~\ref{fig:ZEP-lambda_mean} confirms that this is the case.
Table~\ref{tab.2} lists the parameters used in the simulations; note
that the early onset time conditions are only weakly satisfied ($T'$
ranges from $25 - 40\%$ of $\tau_0$).

\begin{table}[htb]
\begin{center}
\begin{tabular}{|c||c|c|c|c|c|l|l|}
  \hline
  Model & $\tau_0$ & std($\tau_0$) & $T$ &  $\Delta T$ & $p$ & $\lambda^*$ 
  (thy.) & $\lambda^*$ (sim.) \\
  \hline\hline
  CD  & $4.005$ & $0.960$ & $0.5$ & $0.5$ & $5$ & $1$ & $1.0013$ \\
    \hline
  TD  & $3.145$ & $0.420$& $0.5$ &$0.5$ & $5$ & $1$ & $0.9978$ \\
    \hline
  SOU & $1.831$ & $ 0.374$ & $0.1$ & $0.4$ & $2$ & $1.2214$ & $1.2251$  \\
    \hline
  UOU & $2.954$ & $0.142$ & $0.2$ & $1.0$ & $2$ & $0.9048$ & $0.9050$ \\
  \hline
\end{tabular}
\caption{Parameters used in ZEP simulations.  $\tau_0$ and
std($\tau_0$) are the first and second moments of the  passage time
without perturbation,  $T$ and $\Delta T$ are the onset time and total
duration  of the perturbation; $p$ is the amplitude of the second
pulse and $\lambda^*$ (thy.) and $\lambda^*$ (sim.) are the ZEP pulse
ratios predicted by the theory and obtained by averaging $p < 0$ and
$p > 0$ results of Fig.~\ref{fig:ZEP-lambda_mean}. }
\label{tab.2}
\end{center}
\end{table}

To derive explicit approximations for $\lambda^*$, we use a comparison
method similar to that of Appendix A.  Let $X_t$ denote the
unperturbed process and $Z_t$ the perturbed process. If $X_0 = Z_0$ at
$t=0$, then $X_T = Z_T$ at the onset time $T$. When the perturbation
ends at $t = T'$ the solution of the unperturbed SDE of
Eq.~\ref{eq:general} with constant drift rate $b_0$ is
\begin{equation}
X_{T'}= (X_T+b_{0}/k)e^{k\Delta T} - b_{0}/k + \sigma \int_T^{T+\Delta T}
 e^{k(T'-s)} dW_s ,
\label{eq:unpert}
\end{equation}
while the perturbed system satisfies
\begin{equation}
Z_{T'}=(X_T+b_{0}/k)e^{k \Delta T} - b_{0}/k + \int_T^{T + \Delta T}
 e^{k(\Delta T-s)}b_1(s)ds +\sigma \int_T^{T+\Delta T} e^{k(T'-s)} dW_{s} .
\label{eq:pert}
\end{equation}
These expressions well approximate the true activity levels for models
CD, SOU and UOU with absorbing thresholds only when both processes
have low probability of hitting the threshold during the interval
$[0,T']$, but this condition holds for $T'/\tau_0 \ll 1$, since the
integrated drift term is almost zero at early times.
Eqs.~\ref{eq:unpert} and \ref{eq:pert} differ in the term $e^{k \Delta
T} \int_0^{\Delta T} e^{-k t}b_1(t)dt$, which enters $Z_{T'}$  due to
the action of the perturbation $b_1$ during the interval $[T,T']$. The
ZEP condition for arbitrary perturbations is therefore
\begin{equation}
\int_T^{T + \Delta T} e^{-kt}b_1(t)dt = 0 ,
\label{eq:ouzep}
\end{equation}
which for the piecewise constant pulses of Eq.~\ref{eq:ZEP-lambda}
implies that:
\begin{eqnarray}
\lambda^* = \frac{e^{-k\Delta T /2} - e^{-k\Delta T}}{1 - e^{-k\Delta T /2}}
 = e^{-k\Delta T /2} .
\label{eq:ouzep-pwc}
\end{eqnarray}
Hence for $k>0$ (UOU), $\lambda^* < 1$, while for $k<0$
(SOU), $\lambda^* > 1$. In the special case of CD, $k=0$ and
$\lambda^* = 1$. 

For TD the drift term is time-dependent but the linear term
$kX_t$ is absent and so the corresponding solutions of the SDEs are
\[
X_{T'} = \int_T^{T+\Delta T} b_0(s) ds + \sigma \int_T^{T+\Delta T} dW_s
\]
and
\[
Z_{T'} = \int_T^{T+\Delta T} b_0(s) ds + \int_T^{T+\Delta T} b_1(s) ds
 + \sigma \int_T^{T+\Delta T} dW_s .
\]
In this case, a ZEP must satisfy 
\begin{equation}
\int_T^{T + \Delta T} b_1(t) dt = 0 , 
\label{eq:ouzepTD}
\end{equation}
which yields $\lambda^*=1$ for the pulses of Eq.~\ref{eq:ZEP-lambda},
as for CD. This result holds for \emph{all} time-dependent drift rates,
including those used in \citet{Smith2004,Ditterich06b,churchland2008}. 

Summarizing, ZEPs occur for the CD and TD models when the
opposing pulse amplitudes are equal ($\lambda^*=1$), but for SOU and
UOU $\lambda^*=  e^{-k\Delta T/2}$ is larger and smaller than one,
respectively. The ratios $\lambda^*$ predicted by
Eq.~\ref{eq:ouzep-pwc} for the parameters of Table~\ref{tab.2} are
given in the final column of the Table. In all four cases they agree
well with the zero crossings of the mean first passage times in
Fig.~\ref{fig:ZEP-lambda_mean}, in spite of the fact that  $T' / 
\tau_0 \approx 0.25 - 0.4$ is not very small.

\subsection{Signal-to-noise ratio can influence perturbation effects}
\label{ss.SNR}

To what extent do the results of Section~\ref{ss.spplin} hold when
signal-to-noise ratios are reduced and a second threshold is added to
track errors? To investigate this question we select new parameter
values according to the criteria of Section~\ref{ss.sim}. 

\begin{figure}[htb]
 \centering
\includegraphics[scale=0.64]{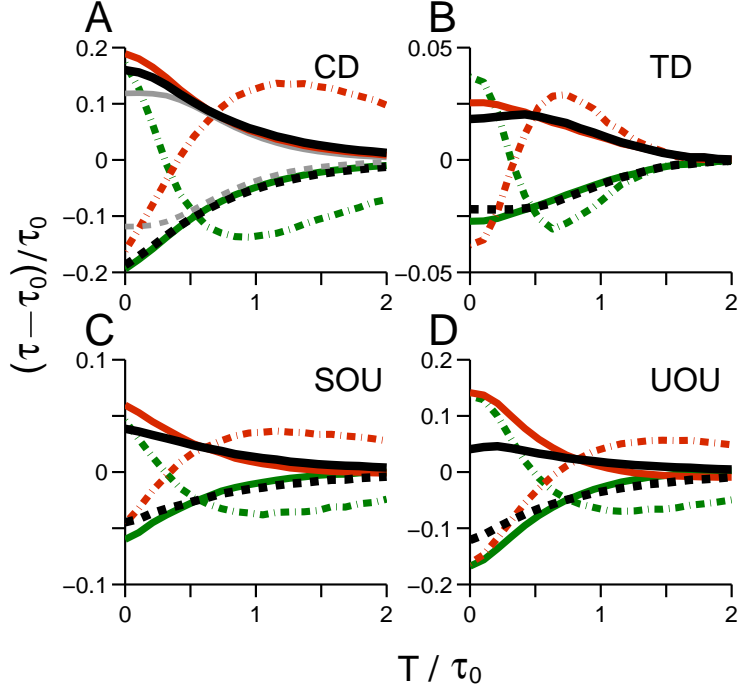}
\caption{Changes in mean decision times for linear models with lower 
signal-to-noise ratios. 
Perturbation protocol is as in Fig.~\ref{fig:T_mean}. Green 
(red) curves: perturbation $p$ in the same (opposite) direction as drift rate
$b_0$; solid: correct choices; dash-dotted: error choices. Solid (dashed) 
black curves are averages of correct and error choices with perturbation 
in the same (opposite) direction as drift rate. Grey (black) curves in panel
A are for error rate of $0.02\%$ ($5\%$). }
\label{fig:noisyRTmean}
\end{figure}

For the CD model, we increase $\sigma$ from $2.449$ to $2.828$ and
set thresholds at $\zeta = \pm 5$ instead of $\pm 20$, yielding
an error rate of $5\%$. This introduces more complex behavior
in which correct and error trials respond to the perturbations in
different manners, as shown in Fig.~\ref{fig:noisyRTmean}A.
The combined averages (as computed in \citet{HukShadlen2005}
and \citet{Wong2007}, see black solid and dashed curves) preserve
some features of previous results: for early perturbations the
direction of changes in mean RT agrees with Fig.~\ref{fig:T_mean}A, 
but the approximately-constant change in mean decision time
early in the trial is replaced by a monotonic decline. Increased noise
yields earlier threshold crossings, advancing the thresholding
effect and masking the signatures of Fig.~\ref{fig:T_mean}A. 
Moreover, the change in mean error
RTs can reverse sign for late perturbation onsets.
We checked this seemingly counterintuitive phenomenon numerically 
by fixing the random generating seed for noise
in a sample trial, finding that perturbation of long RT trials
can change an impending error to a correct choice (data not shown).

Increasing the noise level to $\sigma = 7.071$ with thresholds
$\zeta = \pm 20$ produces an error rate of $10\%$ in the TD model and
changes in mean RTs for correct and error choices similar to 
those of CD. However, averaged over correct and 
error choices, overall changes in mean RT exhibit a pattern similar
to that of Fig.~\ref{fig:T_mean}B (see black curves in
Fig.~\ref{fig:noisyRTmean}B). In this respect TD maintains its
signature more robustly with increasing error rate than CD,
although the effects are progressively masked.

For the SOU model, we increase $\sigma$ to $6.325$, yielding an error
rate of $17\%$. The optimal perturbation onset time for changes in
mean RT averaged over correct and error trials is masked  (compare
Fig.~\ref{fig:T_mean}C with black curves in
Fig.~\ref{fig:noisyRTmean}C). This error rate is relatively high, but
we have checked that for error rates as small as $3\%$ SOU model
features can also be masked, indicating that, like the CD model, SOU
is sensitive to noise.

For the UOU model, more parameters must be adjusted
to obtain reasonable RT changes. Here we set $\sigma=2$,
$\zeta=\pm 10$, $k=0.02$ and $b_0=0.5$, yielding a high error rate 
of $23.5\%$, but similar trends occur for error rates from $0.1\%$ 
to $35\%$. Fig.~\ref{fig:noisyRTmean}D shows that these are comparable
to the previous UOU results with high signal-to-noise ratio. Overall,
we are unable to distinguish between noise-masking effects and
intrinsic features of the UOU integrator.

\begin{figure}[htb]
 \centering
\includegraphics[scale=0.64]{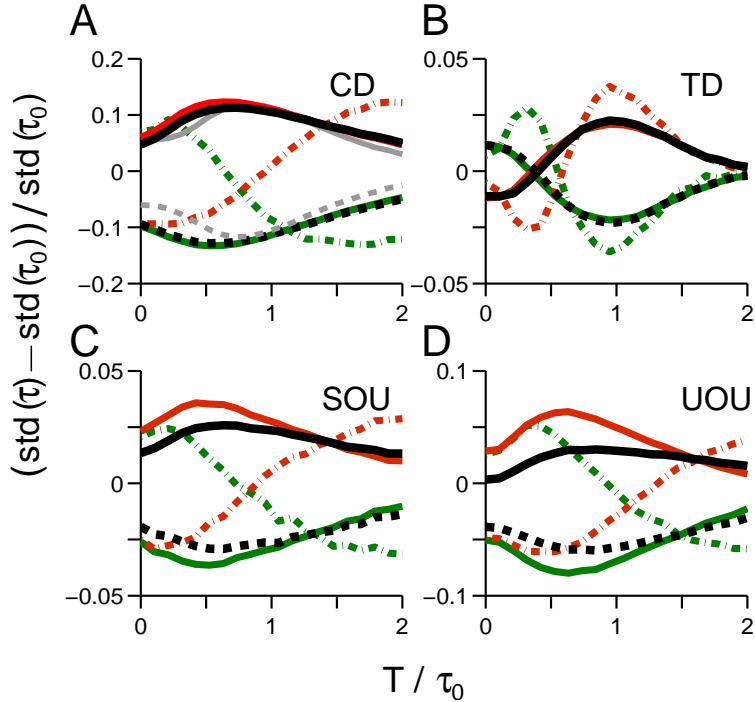}
\caption{Changes in standard deviations of decision times for the
linear models with lower signal-to-noise ratios. 
Format is as in Figs.~\ref{fig:T_std}
and \ref{fig:noisyRTmean}.}
\label{fig:noisyRTstd}
\end{figure}

Using the same parameters, we find that changes
in standard deviations for all the linear models show optimal 
perturbation onset times (Fig.~\ref{fig:noisyRTstd}). The CD, SOU and 
UOU models are mutually indistinguishable (cf. 
Figs.~\ref{fig:noisyRTstd}A, C and D), but TD exhibits a crossing effect 
(Fig.~\ref{fig:noisyRTstd}B) similar to Fig.~\ref{fig:T_std}B.

\subsection{Perturbations of the nonlinear model}
\label{ss.pertnl}

\begin{figure}[htb!]
 \centering
 \includegraphics[scale=0.6]{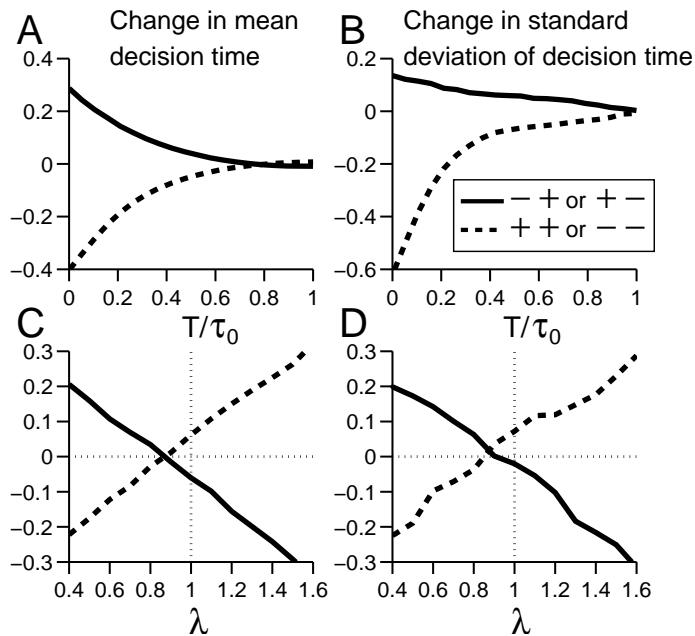}
\caption{Perturbations of means (left column) and standard deviations
(right column) of first passage times for the SPB model. Panels A, B:
effects of single pulse perturbations with varying onset times;
vertical axes as in Figs.~\ref{fig:T_mean} and \ref{fig:T_std}. $- +$
denotes effect of negative pulse at positive threshold, $+ -$ denotes
effect of positive pulse at negative threshold, etc. Due to symmetry,
$- +$ and $+ -$ have same effect (solid curves) as do $+ +$ and $- -$
(dashed curves). Panels C, D: responses to zero-effect perturbations;
vertical axis in C as in Fig.~\ref{fig:ZEP-lambda_mean}; vertical axis
in D shows normalized change in standard deviation as in
Fig.~\ref{fig:T_std}.}
\label{fig:T-ZEP-mean-std-SPB}
\end{figure}

We next apply both perturbation protocols to the nonlinear SPB
model. The mean and standard deviation of decision times  for the
unperturbed system are $80$ and $22$ time units,  respectively. The
parameters used for the first protocol are given in
Table~\ref{tab.1}, while for the ZEP we set $p=0.005$,
$T=30$ and $\Delta T=5$,  consistent with application early in the
accumulation process.  Figs.~\ref{fig:T-ZEP-mean-std-SPB}A and B show
that the means and standard deviations of decision times behave like
those for UOU (cf. panels D of Figs.~\ref{fig:T_mean} and
\ref{fig:T_std}), decreasing monotonically and approaching zero as
onset times increase. ZEPs occur in the pulse ratio range 
$0.8 < \lambda^* < 1$ for both mean and standard deviation
(Figs.~\ref{fig:T-ZEP-mean-std-SPB}C and D). Assuming a linear
change in mean in Fig.~\ref{fig:T-ZEP-mean-std-SPB}C, we find that
$\lambda^* \approx 0.87$. 

\begin{figure}[htb]
 \centering
\includegraphics[scale=0.4]{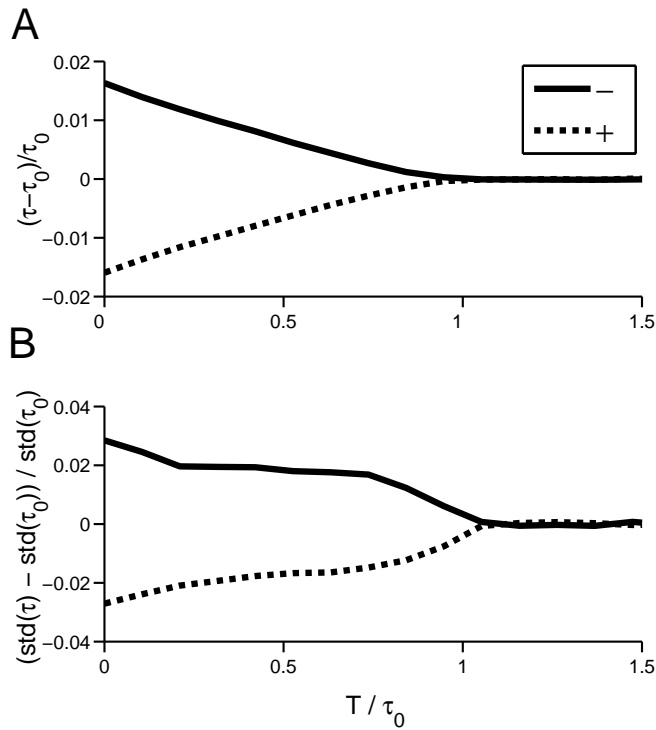}
\caption{Changes in mean (A) and standard deviation (B) of decision
times of the SPB model with high signal-to-noise ratio. All choices
are correct. Perturbation protocols and plot formats are as in 
Figs.~\ref{fig:T-ZEP-mean-std-SPB}A and B. }
\label{fig:SPB_1thres}
\end{figure}

To compare with the behavior of the  linear systems in the single
threshold case, we now modify the SPB model so that it makes as few
errors as they do. This is done by changing $b_0$ from $0$ to $0.004$,
which breaks the  symmetry of the subcritical pitchfork bifurcation
(Fig.~\ref{fig:scheme}B)  so that decisions favor one choice,
representing easy tasks  (cf. \cite{RoxinLedberg08}); all other
parameters remain the same.  Fig.~\ref{fig:SPB_1thres} shows that the
basic behavior of  Figs.~\ref{fig:T-ZEP-mean-std-SPB}C and D is
preserved: effects on both mean and standard deviation of RTs decrease
monotonically as perturbation onset time increases. While the effects
on mean passage times of SPB are much like those for UOU, the fact
that standard deviations of all four linear models exhibit maxima
(Fig.~\ref{fig:T_std}) potentially allows us to distinguish SPB from
them.  The effects of ZEPs in this biased regime is similar to that
shown in Figs.~\ref{fig:T-ZEP-mean-std-SPB}C and D (data not shown).

\section{Discussion}
\label{s.disc}

In this work, we have investigated how several simple stochastic
integrate-to-threshold decision-making models respond to short
perturbations. We focus on two-alternative forced-choice reaction time
tasks for which general firing rate models with three coupled neural
populations can be reduced to one-dimensional integrate-to-threshold
systems. These reduced systems, which may be pure drift-diffusion or 
Ornstein-Uhlenbeck processes, can be compared analytically and in 
simulations that require few parameters. 

We examined two perturbation protocols, the first being a brief pulse
with variable onset time, and the second a pulse-antipulse pair whose
amplitude ratio can be adjusted to produce minimal effects 
(the zero-effect perturbation, ZEP). The
simulations of \S\ref{s.results} (Figs.~\ref{fig:T_mean}
and~\ref{fig:T_std}) show that the changes in means and standard
deviations of decision times allow both perturbation protocols to
distinguish between drift-diffusion models and stable and unstable 
Ornstein-Uhlenbeck processes, provided that perturbations are delivered 
sufficiently early in the integration process and the signal-to-noise 
ratio is not too small. Changes in standard deviations can also
assist in distinguishing drift-diffusion models with constant drift from 
those with time-dependent drift rates (cf. Figs.~\ref{fig:T_std}A and B),
although differences are small and may not be detectable
experimentally. (As \cite{Ditterich2006c} notes, entire
reaction time distributions can further assist in this.) For
early perturbations the ZEP conditions can be approximated
analytically, making predictions that are confirmed by computer
simulations (Fig.~\ref{fig:ZEP-lambda_mean}). While general analytical
expressions for systems with finite noise appear elusive, in 
Appendix A we prove inequalities for more general nonlinear systems 
that partially explain our simulation results. 

Although we focus on linear models, our methods extend to other
systems, and we include results for a nonlinear model that capture the
dynamics near a subcritical pitchfork bifurcation which is typical of
reduced population models
\citep{Wang2002,WongWang2006,RoxinLedberg08}.  As the results of
Appendix A suggest, it behaves in a manner similar to the unstable OU
model under both perturbation protocols,  although perturbative
effects on standard deviations may allow distinctions to be made
(Figs.~\ref{fig:T-ZEP-mean-std-SPB}-\ref{fig:SPB_1thres}). We have
also confirmed that the TD model's behavior remains similar with other
increasing drift rates  (e.g. $b(t) \sim b_{0} t^2$; simulation
results not shown).

These results reinforce the simulations and claims in \citet{Wong2007} 
that time-varying perturbations can in principle reveal integration
mechanisms. However, some parameter ranges frustrate
clear-cut comparisons among models. The first occurs for
low signal-to-noise ratios and high error rates
(cf. Figs~\ref{fig:noisyRTmean}-\ref{fig:noisyRTstd}), and so can be
mitigated by using perturbations with more easily discriminated stimuli. 
Secondly, if there is a `dead time' between stimulus onset
and the start of evidence integration, the CD and TD models can
display similar signatures to the SOU model, even with high
signal-to-noise ratios.

Since perturbations can be delivered through the senses as well as by direct
electrical stimulation, the method can be noninvasive and is therefore
appropriate for human subjects. Furthermore, unlike fitting
neuronal firing rates, both perturbation protocols can
distinguish linear and nonlinear UOU-type models 
\citep{Wang2002,WongWang2006,RoxinLedberg08,WongHuk08} from drift-diffusion 
models with an `urgent' time-dependent increase in drift rate
\citep{Smith2004,Ditterich2006c,churchland2008}. 
However, care is required in designing such experiments. 
Although a high signal-to-noise ratio can in principle help
identify integrators, the reaction times 
can become too short, masking the signature behaviors of the 
integrators with the `thresholding effect'. A possible solution 
could be to prolong the response deadline in relatively simple 
reaction time tasks and analyze only data from long RT trials. 

Perturbation inputs to decision-making circuits may originate in a
single sensory pathway or in other brain areas, as in multisensory and
`top-down' inputs (e.g. due to attention \citep{Smith2004,LiuHCoh08}) 
which could modulate a decision throughout a sensorimotor pathway. 
The general notion of perturbation developed in the reduced model 
context extends to such modulatory inputs. The methods may also
generalize to other tasks that recruit neural integrators
\citep{goldman2008}, such as interval timing \citep{BrownRinzel2006}.


\section*{Appendix A: Rigorous estimates on first passage times}
\appendix

In this appendix we analyze short single pulse perturbations, First
in the noiseless limit, and then with additive
noise. The inequalities proved here suggest explanations for the
monotonicity of mean first passage times of SOU and UOU processes with
respect to early onset times (cf. Figs.~\ref{fig:T_mean}C and D);
moreover, they apply to more general, nonlinear one-dimensional
systems such as SPB  with thresholds $|\zeta|$ sufficiently close to
$X_{t}=0$.

\subsection*{The noiseless limit}

As noise amplitude tends to zero, the SDE Eq.~\ref{eq:general} 
becomes an ordinary differential equation (ODE):
\begin{equation} 
\frac{dX}{dt} := \dot{X} = f(X) .
\label{eq:ode0}
\end{equation}
Consider two compact perturbations $p$ and $\tilde{p}$ with the
same amplitude profile $g(t)$ and duration $\Delta T$, applied at
different onset times $T_1$ and $\tilde{T}_1$:
\begin{equation}
p(t)=1_{[T_1,T_2]}(t)g(t-{T}_1), \quad
\tilde{p}(t)=1_{[\tilde{T}_1,\tilde{T}_2]}(t)g(t-\tilde{T}_1) 
\label{eq:pert1}
\end{equation}
(here $1_{[T_1,T_2]}$ denotes the indicator function that takes the
value 1 for $T_1 < t < T_2$ and zero otherwise). We choose $0 < T_1 <
T_2 < \tilde{T}_1 < \tilde{T}_2$ and $T_2-T_1 = \tilde{T}_2 -
\tilde{T}_1 = \Delta T$ such that $p$ and $\tilde{p}$ do not overlap
and that neither perturbed solution reaches the threshold $X = \zeta$
before $t = \tilde{T}_2$. We shall compare solutions $Y_t$ and
$\tilde{Y}_t$ of the corresponding perturbed ODEs,
\begin{equation}
\dot{Y} = f(Y) + p(t)  \quad  \mbox{and} \quad
\dot{\tilde{Y}} = f(\tilde{Y}) + \tilde{p}(t) ,
\label{eq:ode1}
\end{equation}
started at the same initial condition $Y_0=\tilde{Y}_0=0$, to show how
the signs of $f'$ and $g$ determine their threshold passage times
$\tau$ and $\tilde{\tau}$. We allow smooth nonlinear functions $f$,
but assume that $f(X) > 0$ for all $0 \le X < \zeta$ and that $g$ does not
change sign. The fact that solutions of scalar ODEs cannot cross each
other is a key tool.


\vspace{0.2cm} \noindent
\textbf{Lemma 1:} \emph{Let $X_t(x)$ denote the solution of
Eq.~\ref{eq:ode0} with initial value $x$. Then $X_t(x_1) < X_t(x_2)$
for any $t>0$ if and only if $x_1 < x_2$.}

\vspace{0.2cm} \noindent 
{\bf Proof:} This follows from the uniqueness of solutions
of ODEs and the ordering of points in a one-dimensional phase space.
\hfill $\Box$ 
\vspace{0.1cm}

Since neither perturbation acts after $t = \tilde{T}_2$, and neither
solution has reached threshold at $\tilde{T}_2$, Lemma 1 implies that
$\tau < \tilde{\tau}$ if and only if  $Y(\tilde{T}_2) >
\tilde{Y}(\tilde{T}_2)$. We now prove the main result summarized in
Table~\ref{tab:appthem}, which is a corollary of Proposition 2.

\begin{table}[htb]
  \centering
\begin{tabular}{|c|c|c|}
\hline
 & $f'\geq L >0$ & $f' \leq -L<0$ \\ \hline\hline
$g>0$&$ \tau<\tilde{\tau}$& $ \tau > \tilde{\tau}$  \\ \hline
$g<0$&  $\tau>\tilde{\tau}$& $\tau<\tilde{\tau}$
\\
\hline
\end{tabular}
\caption{Monotonicity of first passage dependence on 
onset time for four sign combinations of $f^\prime$ and $g$, in
the noise-free limit. $L$ is some constant.} 
\label{tab:appthem}
\end{table}


\vspace{0.2cm} \noindent
\textbf{Proposition 2:} \emph{Assume that the function $g(t)$ in
Eqs.~\ref{eq:pert1} is strictly positive for $t \in (0,\Delta T)$. Then 
for some constant $L$ (a) If  $f' \geq L > 0$, then $Y(t) > \tilde{Y}(t)$ 
for all $t\geq \tilde{T}_2$; (b) If $f' \leq -L < 0$, then 
$Y(t) < \tilde{Y}(t)$ for all $t\geq \tilde{T}_2$.}

\vspace{0.2cm} \noindent 
{\bf Proof of part (a):}  We need only establish the
inequalities at $t=\tilde{T}_2$, since no perturbations occur after
this time. We compare solutions of Eqs.~\ref{eq:ode1} using the fact
that $f(X) > 0$ and $f'(X) \geq L > 0$ for $X \ge 0$ implies that $f(Y)
- f(\tilde{Y}) \ge L(Y - \tilde{Y})$. Letting $u(t) = Y(t) -
\tilde{Y}(t)$ we have
\begin{equation}
\dot{u} = f(Y) - f(\tilde{Y}) + p(t) - \tilde{p}(t) \ge L u
 + p(t) - \tilde{p}(t) .
\label{eq:ode2}
\end{equation}
Since the solutions coincide until the first perturbation occurs
at $t = T_1$, $u(T_1) = 0$, and integrating inequality 
\ref{eq:ode2} over the interval $[T_1, T_2]$, on which $p(t) =
g(t - T_1)$ and $\tilde{p}(t) \equiv 0$, yields:
\begin{equation}
u(T_2) \ge \int_{T_1}^{T_2} e^{L(T_2-t)} g(t-T_1) dt > 0 .
\label{eq:ode3}
\end{equation}
Both $p(t)$ and $\tilde{p}(t)
\equiv 0$ on $[T_2, \tilde{T}_1]$, and integration of \ref{eq:ode2}
and use of \ref{eq:ode3} gives
\begin{equation}
u(\tilde{T}_1) \ge u(T_2) e^{L(\tilde{T}_1-T_2)} \ge 
e^{L(\tilde{T}_1-T_2)} \int_{T_1}^{T_2} e^{L(T_2-t)} g(t-T_1) dt > 0 .
\label{eq:ode4}
\end{equation}
Finally, using the fact that $p(t) \equiv 0$ on $[\tilde{T}_1, 
\tilde{T}_2]$, we integrate \ref{eq:ode2} again to obtain
\begin{eqnarray}
u(\tilde{T}_2) & \ge & u(\tilde{T}_1) e^{L(\tilde{T}_2 - \tilde{T}_1)} 
- \int_{\tilde{T}_1}^{\tilde{T}_2} e^{L(\tilde{T}_2-t)} g(t-\tilde{T}_1) dt
\nonumber \\
& \ge & \left[ e^{L(\tilde{T}_2-T_2)} - 1 \right] \int_{0}^{\Delta T}
 e^{L(\Delta T - t)} g(t) dt > 0 ,
\label{eq:ode5}
\end{eqnarray}
where we also use inequality \ref{eq:ode4} and the fact that the
additive perturbations of Eq.~\ref{eq:pert1} are identical over the
time intervals $[T_1, T_2]$ and  $[\tilde{T}_1, \tilde{T}_2]$ to
rewrite the integral  term. Hence $Y(\tilde{T}_2) >
\tilde{Y}(\tilde{T}_2)$ and $\tau < \tilde{\tau}$.

\vspace{0.1cm}

\noindent {\bf Proof of part (b):}  We again compare solutions of
Eqs.~\ref{eq:ode1} but the direction of the inequalities is now
reversed, so that we are in essence using Gronwall's inequality 
\citep{GH83}. In this case Eq.~\ref{eq:ode2} becomes
\begin{equation}
\dot{u} \le -L u + p(t) - \tilde{p}(t) ,
\label{eq:ode6}
\end{equation}
inequality~\ref{eq:ode4} becomes
\begin{equation}
u(\tilde{T}_1) \le u(T_2) e^{-L(\tilde{T}_1-T_2)} \le 
e^{-L(\tilde{T}_1-T_2)} \int_{T_1}^{T_2} e^{-L(T_2-t)} g(t-T_1) dt < 0 ,
\label{eq:ode7}
\end{equation}
and the final inequality reads
\begin{equation}
u(\tilde{T}_2) = Y(\tilde{T}_2) - \tilde{Y}(\tilde{T}_2)
\le - \left[ 1 - e^{-L(\tilde{T}_2-T_2)} \right] 
\int_{0}^{\Delta T} e^{-L(\Delta T - t)} g(t) dt < 0 .
\label{eq:ode8}
\end{equation}
This implies that $\tau > \tilde{\tau}$, as claimed. Proofs are similar
for $g < 0$.  \hfill $\Box$ \vspace{0.3cm}

\subsection*{Extension to noisy systems}

The conclusions of Proposition~2 and Table~\ref{tab:appthem} may be extended to
apply to expected passage times for SDEs with additive noise. 
Specifically, replacing Eqns.~\ref{eq:ode1} by 
\begin{subequations}
\begin{align}
d Y_{t} = [f(Y_t) + p(t)] dt + \sigma dW_{t} ,
\label{eq:sode1a} \\
d \tilde{Y}_{t} = [f(\tilde{Y}_t) + \tilde{p}(t)] dt + \sigma d \tilde{W}_{t} ,
\label{eq:sode1b}
\end{align}
\label{eq:sode1}
\end{subequations}
we have the following result:

\vspace{0.2cm} \noindent
\textbf{Proposition 3:} \emph{Assume that the function $g(t)$ in
Eqs.~\ref{eq:pert1} is strictly positive for $t \in (0,\Delta T)$. Then 
for some constant $L$ (a) 
If  $f' \geq L > 0$, then $\mathbb{E}[\tau] < \mathbb{E}[\tilde{\tau}]$;
(b) If $f' \leq -L < 0$, then $\mathbb{E}[\tau] > \mathbb{E}[\tilde{\tau}]$,
where $\tau$ and $\tilde{\tau}$ denote the first passage times for 
Eqns.~\ref{eq:sode1a} and \ref{eq:sode1b} respectively.}

\vspace{0.2cm} \noindent 
{\bf Proof:}  We first compare solutions $Y_{t}$ and $\tilde{Y}_{t}$
obtained from Eqns.~\ref{eq:sode1a} and \ref{eq:sode1b} under the
\emph{same} sample noise path with increments $d \tilde{W}_{t} \equiv
dW_{t}$. Since the stochastic integral $\int_0^t dW_{t}$ is almost
surely continuous \citep{Feller57}, and the 
integrated noise terms cancel, the comparison of solutions using $u(t)
= Y_t - \tilde{Y}_{t}$ proceeds as in the deterministic  case above,
and we conclude that $\tau < \tilde{\tau}$ for $g > 0$ and $f'\geq L >
0$, and $\tau > \tilde{\tau}$ for $g > 0$ and $f' \leq -L <
0$. Finally, averaging each process over an ensemble of sample paths,
we may conclude that the mean first passage times of the two processes
satisfy similar  inequalities to those in the top row of
Table~\ref{tab:appthem}:  $\mathbb{E}[\tau] <
\mathbb{E}[\tilde{\tau}]$ and $\mathbb{E}[\tau] >
\mathbb{E}[\tilde{\tau}]$ respectively. A similar argument applies
to the case $g < 0$.   \hfill $\Box$ 


\section*{Appendix B: A TD model with time-varying perturbation amplitude}

\begin{figure}[htb]
 \centering
\includegraphics[scale=0.45]{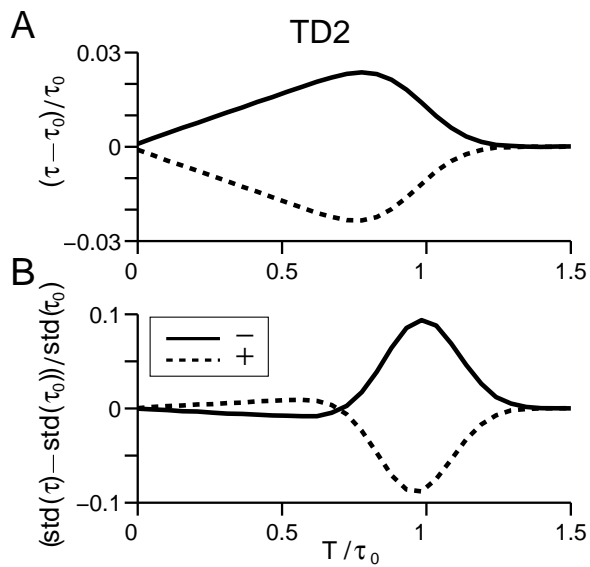}
\caption{Normalized changes in mean (A) and standard deviation (B) of 
first passage times for single pulse perturbations with varying onset 
times applied to a variant of the TD model. Vertical axes and curve
conventions are as in Figs.~\ref{fig:T_mean} and \ref{fig:T_std}. }
\label{fig:TD2}
\end{figure}

In the body of the paper, perturbations of the TD model were assumed 
unaffected by the time-dependent gain. Here we extend the simulations
to a simple case in which perturbation and stimulus (drift) share the same
overall linear increase in amplitude. Specifically, we assume the 
dynamics to be governed by 
\begin{eqnarray}
d X_{t} = (b_{0}+ p(t))t dt + \sigma dW_{t} .
\label{eq:TD2}
\end{eqnarray}
Otherwise, we change only the perturbation amplitude, taking
$p = \pm 2$ instead of $4$. 
Fig.~\ref{fig:TD2}A shows that the changes in mean and standard 
deviations of exit time for this model remain similar to 
the previous TD model, but with smaller perturbative effects 
near the beginning of the trial (quite similar to a SOU model) due to the 
monotonic time-dependence of perturbation amplitude. However,
although the previous pattern of mean exit times no longer applies,
the behavior remains qualitatively \emph{dissimilar} to the CD, 
UOU and SPB models.


\vspace{0.5cm} \noindent {\bf{Acknowledgments}} \
We thank C.~D.~Brody for suggesting a variant of the TD model,
M. ~Usher and an anonymous reviewer for helpful suggestions
that improved the paper, and Nengli Lim for pointing out the 
extension to noisy systems in Proposition~$3$ of Appendix A. 
XZ was supported in part by ONR grant N00014-01-1-0674, and KFWL and PH
by PHS grants MH58480 and MH62196 and AFOSR grant FA9550-07-1-0537. 
The U.S. Government is authorized to reproduce and distribute reprints 
for Governmental purposes notwithstanding any copyright notation
thereon. The views and conclusions contained herein are those of the
authors and should not be interpreted as necessarily representing the
official policies or endorsements, either expressed or implied, of the
Air Force Research Laboratory or the U.S. Government.


\bibliography{timepertint}

\end{document}